# CONTROL OF PHOTON TRANSPORT PROPERTIES IN NANOCOMPOSITE NANOWIRES

M. Moffa[a], V. Fasano[b], A. Camposeo[a], L. Persano[a], D. Pisignano*[a,b]

[a]Istituto Nanoscienze-CNR, Euromediterranean Center for Nanomaterial Modelling and Technology (ECMT), via Arnesano, I-73100 Lecce (Italy); [b]Dipartimento di Matematica e Fisica "Ennio De Giorgi", Università del Salento, via Arnesano, I-73100 Lecce (Italy).

## ABSTRACT

Active nanowires and nanofibers can be realized by the electric-field induced stretching of polymer solutions with sufficient molecular entanglements. The resulting nanomaterials are attracting an increasing attention in view of their application in a wide variety of fields, including optoelectronics, photonics, energy harvesting, nanoelectronics, and microelectromechanical systems. Realizing nanocomposite nanofibers is especially interesting in this respect. In particular, methods suitable for embedding inorganic nanocrystals in electrified jets and then in active fiber systems allow for controlling light-scattering and refractive index properties in the realized fibrous materials. We here report on the design, realization, and morphological and spectroscopic characterization of new species of active, composite nanowires and nanofibers for nanophotonics. We focus on the properties of light-confinement and photon transport along the nanowire longitudinal axis, and on how these depend on nanoparticle incorporation. Optical losses mechanisms and their influence on device design and performances are also presented and discussed.

## 1. INTRODUCTION

Active, polymer-based nanowires and nanofibers exhibit very interesting optical and electronic properties [1, 2]. These structures consist of nanorods or filaments, with diameter ranging from a few nm up to roughly 1 μm, and they can be produced by a variety of methods which include self-assembly, in turn often promoted by polymer precipitation involving marginal solvents, the use of either hard or soft templates, polymerization techniques, nanofluidics and soft lithographies [3,4]. However, it is through so-called electrostatic spinning or electro-spinning [5-7] that functional versatility and good production rates combine together to make the production of polymer nanofibers appealing for the largest variety of application fields. Electrospinning technologies exploit an intense electric field applied to a polymer solution, which has to feature a sufficient amount of molecular entanglements, namely good enough viscoelastic behavior. Applied voltages are in kV-range, over distances from a few mm to tens of cm. Near-field methods using mm or sub-mm inter-electrode distance have been also proposed, suppressing instabilities and allowing the position of single wires to be better controlled [8,9]. Regardless of the specific electrospinning approach, polymer solutions are prepared, and injected through a needle biased by one of the two terminals of a voltage supply. An electrified jet is formed in this way, carrying the polymer from the needle to a collector, and continuously reducing its diameter along its path due to the concomitant acceleration of the fluid body and the rapid evaporation of the solvent component. This leads to the deposition of nanofibers onto the collecting surface, and both isolated nanostructures and dense mats can be produced depending on the collector geometry and motion. The properties of the resulting materials critically depend on the chemical species involved, including the used solvents, as well as on the morphology obtained both at the scale of single filaments and in complex mats made of many fibers. For instance, the higher molecular order which can be eventually primed by the strong stretching and elongation of the jet might lead to nanofibers with polymer chains prevalently oriented along the longitudinal axis of the wires, hence to improved charge-carrier transport, polarized emission, anisotropic crystalline structure which in turn can induce stable piezoelectric behavior, and enhanced mechanical behavior [10-13]. Examples of devices which may benefit from such properties include field-effect transistors [9,14], luministors [15], light-emitting devices [16-18], and flexible nanogenerators [19].

A further degree of freedom in terms of material composition which allows physical properties to be tailored can be given by the addition of nanoparticles of various nature to the solutions to be electrospun [20]. In this way, hybrid fibers can be realized, which incorporate inorganic nanocrystals or other species of nanoparticles in the polymer matrices, thus exhibiting new functionalities as provided by the embedded component. Reported examples of hybrid electrospun nanofibers with incorporated particles are numerous, and deal with fillers of different shape, nature, and crystallinity. For instance, single-walled carbon nanotubes have been included in electrospun fibers to produce reinforced yarns [21]. Fluorescent CdTe quantum dots (QDs) have been dispersed into poly(vinyl alcohol) (PVA) fibers, leading to composites with suppressed Förster energy transfer between the particles [22]. Nanodiamonds have been loaded at high concentrations in polyacrylonitrile and in polyamide, leading to hybrid coatings which are interesting for UV protection and scratch resistance [23]. Au nanorods have been aligned along PVA nanofibers, thus generating flexible substrates for surface-enhanced Raman scattering [24]. Functionalized Au nanoparticles in poly(3-hexylthiophene) filaments have been used to realize flexible nonvolatile flash memory devices [25], and so on.

*dario.pisignano@unisalento.it; phone +39 0832298104; fax +39 0832298146; www.nanojets.eu

Furthermore, which is more interesting for designing applications in nanophotonics, the polymer component provides the resulting material with a topological, possibly three-dimensional and free-standing network made by the filaments, in which nanoparticles can constitute distributed functional domains, namely a spatially-resolved information embedded in the fibrous structures. The inter-particle interactions are then mediated by the polymer component, and the resulting spatial modulation of physical properties in the hybrid material can involve refractive index ($n$), due to the contrast between the polymer and the fillers, emission or optical gain properties, and light-scattering characteristics of the whole system. Here we report on the production and characterization of nanocomposite nanowires embedding different classes of nanoparticles, designed to vary either the local refractive index or the emission characteristics. Bright photoluminescence is obtained by CdSe nanoparticles embedded in poly(methyl methacrylate) (PMMA) nanofibers, exhibiting waveguiding capability of self-emitted light along the length of fibers, and optical losses around 200 cm$^{-1}$. More in general, the inclusion of nanoparticles allows the light-scattering properties affecting optical losses to be varied, through the changes induced in the fiber surface roughness. While these aspects might lead to higher optical losses dealing with photon transport along the longitudinal axis of nanocomposite fibers, the capability of tailoring light-scattering properties can be effectively exploited to design new materials for disordered photonics.

## 2. METHODS

All the pristine materials used in this work are commercially available. Electrospinning is performed by preparing chloroform solutions of PMMA (molecular weight = 120,000 Da) with polymer concentrations in the range 150-375 mg/mL and adding nanoparticles, filling a syringe pump (Harvard Apparatus), and using a high-voltage power supply (Gamma High Voltage Research) to apply a voltage bias of 9-16 kV. Different classes of nanoparticles are used as fillers for the nanocomposite wires, namely $TiO_2$ nanoparticles with average diameter 21 nm (Sigma-Aldrich), and light-emitting CdSe QDs with diameter of either 3.4 nm or 6.3 nm (Sigma-Aldrich). The typical needle-collector distances are of 15-20 cm, and steady flows of 0.5-1 mL/h are provided by the pump. An organic salt, tetrabutylammonium iodide is added to the solutions containing the $TiO_2$ nanoparticles, with concentration 15 mg/mL. Scanning electron microscopy (SEM) is carried out by using a FEI Nova NanoSEM 450 system, with 5 kV acceleration voltage. The morphology of the filaments and their surface topography is investigated by atomic force microscopy (AFM). To this aim, a multimode head (Veeco) is used in combination with a Nanoscope IIIa electronic controller. The fiber surface is imaged in tapping mode by using Si cantilevers. The realized nanofibers are also imaged by confocal microscopy, using a laser-scanning microscope relying on an inverted microscope (Eclipse Ti, Nikon) and a spectral scan head (Nikon). The emission of PMMA/CdSe nanowires is excited by a diode laser ($\lambda_{exc}$=408 nm) and measured by a spectral detection unit equipped with a multi-anode photomultiplier. Photoluminescence measurements of light-emitting nanocomposite nanofibers are performed by exciting samples with a continuous-wave diode laser ($\lambda_{exc}$=405 nm, μLS Micro Laser Systems, Inc.), and by collecting the emission through a long-pass filter with cutoff wavelength 450 nm (Thorlabs) and a fiber-coupled spectrometer (USB 4000, Ocean Optics). Photon transport along the longitudinal axis of individual nanocomposite nanowires are investigated according to the micro-photoluminescence method reported in Ref. [26]. Briefly, the above diode laser is used as excitation source to pump a small region along the body of individual light-emitting filaments, through a 20× microscope objective with numerical aperture = 0.5. Self-emitted light generated in the nanowire then propagates along its longitudinal axis, reaching the fiber tip. The intensity of light exiting the fiber body and the tip is then measured by a Peltier cooled charge-coupled device camera (CCD DFC 490, Leica), providing two-dimensional maps of the scattered light, and optical losses are estimated by determining the dependence of collected light on the distance, $d$, between the excited and the observed regions. Absorption propagation losses are estimated by measuring the transmission spectra of a reference film by using a spectrophotometer (Perkin Elmer). The samples are made by spin-casting the solution used for the electrospinning processes onto quartz substrates. The thickness of the used films is measured by using a stylus profilometer.

## 3. RESULTS AND DISCUSSION

Figure 1a-c shows photographs of various species of electrospun nanocomposite samples, composed by PMMA nanowires doped with $TiO_2$ (a), or with luminescent CdSe nanoparticles (b, c). Issues related to the realization of hybrid nanofibers incorporating inorganic fillers include the possible aggregation of particles in solution and the influence on the viscoelastic properties of the electrospun solutions, and the eventual clogging of the metallic needle at the spinneret. However, properly combining the concentration of the spun solutions and the process parameters (mainly injection flux and applied electric field) generally allows reology and clogging issues to be overcome, and stable jets to be achieved. An example of resulting PMMA filaments embedding $TiO_2$ nanoparticles is shown in the SEM micrographs in Fig. 1d,e. $TiO_2$ fillers have a refractive index remarkably higher than PMMA, which can be used to design fibers with distributed variations of $n$ along their length. The here realized nanofibers have average diameter of 340 nm, and a significant surface roughness which is induced by the incorporated fillers. Nanoparticles can be distinguished in SEM micrographs as small protrusions along the fibers (Fig. 1e), and the resulting corrugations on the body of individual nanowires are of the order of tens to hundreds of nm. The overall morphology of PMMA/CdSe nanocomposite fibers is instead inspected by confocal microscopy, showing bright and uniform emission, which suggests a quite uniform dispersion of the inorganic component in the polymer matrix (Fig. 1f,g). From their fillers, nanofibers inherit the emission stability as well as the spectral dependence of the photoluminescence on the size of embedded QDs, which is useful for straightforwardly controlling the emitted color as shown in Figure 2a.

Waveguiding measurements at the scale of single fibers are also carried out, since they provide useful information about the photon transport mechanisms in the nanocomposite filaments. On one side, the dark fiber bodies in the micro-photoluminescence maps displayed in Fig. 2b are indicative of effective light confinement in the volume of the wire.

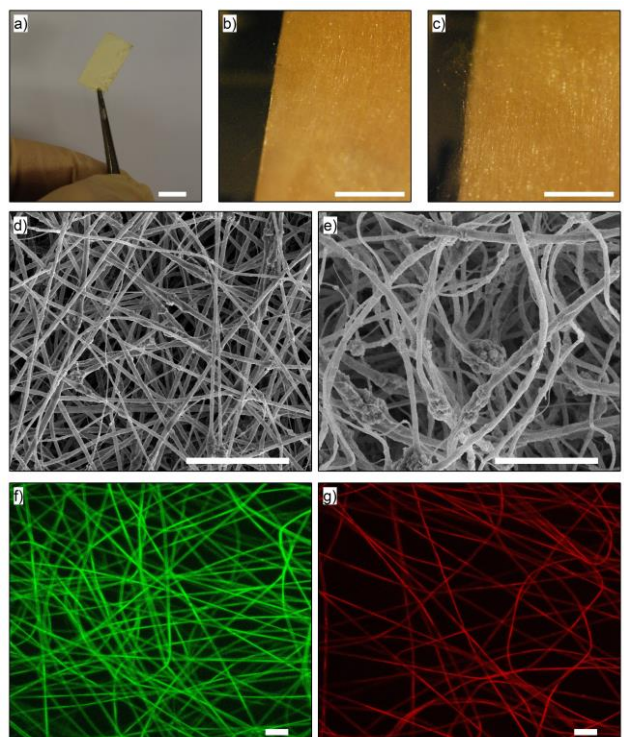

Figure 1. (a): Photograph of electrospun, PMMA/$TiO_2$ nanocomposite mats. Scale bar = 5 mm. (b, c) Photographs of electrospun PMMA fibers mats doped with CdSe QDs having different diameters, 3.4 nm (b) and 6.3 nm (c), respectively. The bright and the dark areas in the photographs correspond to regions covered by the light-emitting nanofibers and to the sample background, respectively. Scale bar = 3 mm. (d, e) SEM micrographs of PMMA/$TiO_2$ nanocomposite nanofibers, imaged at different magnifications. Scale bar = 10 μm (d) and 5 μm (e), respectively. (f, g) Fluorescence confocal micrographs of PMMA fibers doped with CdSe QDs of different diameters, 3.4 nm (f) and 6.3 nm (g). Excitation wavelength = 408 nm. Scale bars = 10 μm.

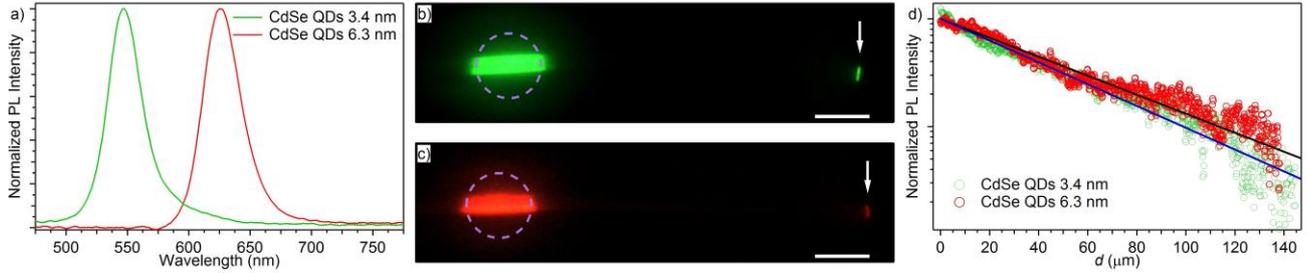

Figure 2. (a) Normalized photoluminescence spectra of PMMA electrospun fibers doped with CdSe QDs of different diameters, 3.4 nm (green line) and 6.3 nm (red line), respectively. (b, c) Micrographs of individual light-emitting PMMA/CdSe nanowires excited by a focused laser spot (violet circles). White arrows in (b) and (c) indicate the fiber tips. Excitation wavelength = 405 nm. Scale bars = 20 μm. (d) Spatial decays of the self-emitted light intensity waveguided along single fibers vs distance, $d$, from the excitation spot. Green circles: PMMA fiber doped with 3.4 nm CdSe QDs. Red circles: PMMA fiber doped with 6.3 CdSe QDs. Continuous lines: best fits to exponential decays (Eq. 1).

Side-losses are however present, and favored in case of rough fibers as those produced with larger or partially aggregated nanoparticles (Fig. 1e). For PMMA-CdSe nanowires, the decrease of the light intensity along their length, corresponding to the distance, $d$, to the excited region, is shown in Fig. 2c. Data are well-fitted by the law:

$$I = I_0 \exp(-\alpha d), \tag{1}$$

where $I$ and $I_0$ indicate the light intensity measured at a given point along the fiber and the light intensity at the excitation point, respectively, and $\alpha$ is the optical loss coefficient. For the two species of PMMA-CdSe nanowires under study, optical losses are comparable (200-230 cm$^{-1}$, corresponding to 870-1000 dB cm$^{-1}$). The estimated light transport length ($l_{tr}=1/\alpha$) is of the order of 50 μm.

These findings are in agreement with the analogous values of the Stokes shifts (50 meV) of the CdSe QDs with either 3.4 nm or 6.3 nm diameter, leading to comparable self-absorption contributions to optical losses. The loss coefficient, $\alpha$, for light propagating along a dielectric filament depends on various concurrent sources:

$$\alpha = \alpha_{Abs} + \alpha_{Rs} + \alpha_{Rss+} \alpha_{Sub}, \tag{2}$$

where $\alpha_{Abs}$ is the contribution due to the absorption of the polymer matrix and of the embedded fillers, $\alpha_{Rs}$ is due to Rayleigh scattering by bulk defects, mainly related to the refractive index contrast between the polymer and the inorganic component, $\alpha_{Rss}$ is related to the Rayleigh scattering by the fiber surface defects, and $\alpha_{Sub}$ is the contribution from light propagating in the filaments which may couple with the substrate underneath. In this study some contributions can be neglected. For instance, the absorption of the polymer matrix (PMMA) is of the order of 10$^{-3}$-10$^{-2}$ cm$^{-1}$ in the visible range [27]. Similarly, the coupling between light propagating in an electrospun PMMA filament and the underlying quartz substrate is not relevant for fibers having micrometer diameter ($D$), whereas this component can become significant upon decreasing the fiber size below the wavelength of the propagating light [26], namely for filaments with subwavelength transversal size. The contribution due to absorption from QDs and from Rayleigh scattering from bulk defects ($\alpha_{Bulk} = \alpha_{Abs} + \alpha_{Rs}$) can be determined by measuring the losses of an optical beam propagating through a film of known thickness, having a composition similar to that of electrospun fibers. Figure 3a shows an example of the spectrum of losses for visible light passing through a film of PMMA doped with CdSe QDs with either 3.4 nm or 6.3 nm diameter. These data allow $\alpha_{Bulk}$ at the emission peak wavelengths (547 nm and 626 nm for 3.4 nm and 6.3 nm CdSe QDs respectively) to be estimated as 50 cm$^{-1}$ and 25 cm$^{-1}$, respectively. Such values evidence that both absorption and scattering from the fillers play a major role in limiting light propagation, as observed also for dye-doped and conjugated polymer fibers [2, 26]. Finally, Rayleigh scattering from surface defects can also be a significant source of optical losses. In order to obtain an estimate of the order of magnitude of such losses one can use an expression based on the Rayleigh criterion and originally derived for slab waveguides [28]:

$$\alpha_{Rss} = \left(\frac{4\pi}{\lambda}\right)^2 \left(\frac{\cos^3\theta}{2\sin\theta}\right)\left(\frac{\sigma^2}{D+l_p}\right). \tag{3}$$

In the above equation, $\theta$ is the angle of the propagating mode with wavelength $\lambda$ ($\theta > 75°$ for PMMA waveguides), $l_p$ is the mode penetration depth into the surrounding medium ($l_p \ll D$ for fibers with micrometric size [29]) and $\sigma$ is the variance of the fiber surface roughness. The characteristics of the fiber surface can be determined by AFM analysis, providing a value $\sigma = 3$ nm for PMMA fibers (Fig. 3b-d), which evidences the good surface uniformity of electrospun polymer fibers. The variance of the fiber surface roughness increases to few tens of nanometers for fibers doped with inorganic fillers, depending on the size and distribution of the embedded nanoparticles. Based on the measured values of $\sigma$, we expect a value for $\alpha_{Rss}$ in the range 1-100 cm$^{-1}$. This analysis shows that the main sources of losses affecting photon transport along the length of electrospun fibers made of a transparent polymer doped with inorganic particles are the absorption and Rayleigh scattering by the embedded particles, and the Rayleigh scattering by surface defects. The former are largely related to intrinsic geometrical and optical properties of the embedded particles, consequently they can be controlled to some extent by varying the concentration and dispersion of fillers in the polymer matrix. Instead, surface scattering can be decreased by improving the fiber uniformity, which can be achieved by a tighter control of the solution properties and of the environmental parameters affecting the electrospinning process.

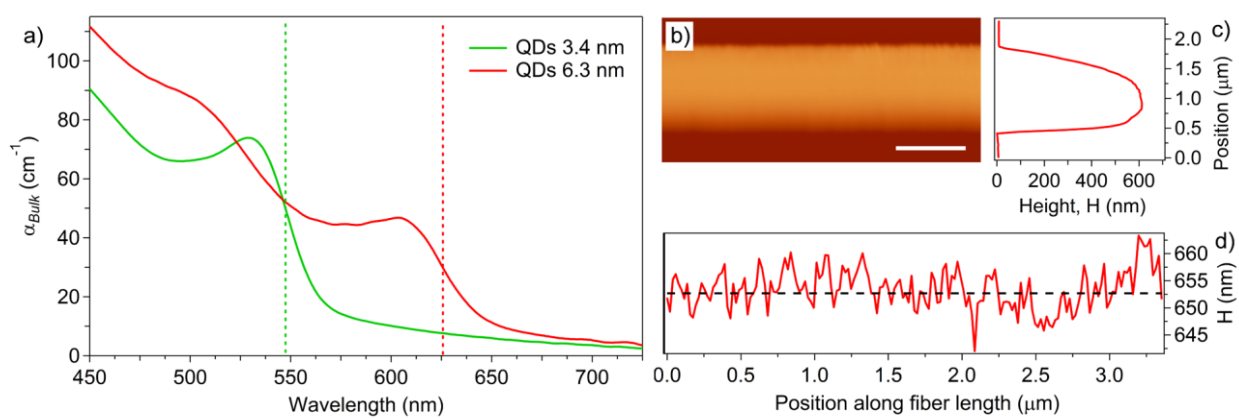

Figure 3. (a) Wavelength dependence of the bulk propagation losses, $\alpha_{Bulk}$, of PMMA doped with CdSe QDs of different diameters, 3.4 nm (green line) and 6.3 nm (red line), respectively. The dashed vertical lines highlight the peak emission wavelength for each sample. (b) AFM micrograph and (c) corresponding height profile of an electrospun PMMA fiber. Scale bar = 1 μm. (d) Variation of the fiber height, H, along the fiber length. The horizontal dashed line marks the average height value (647 nm).

## 4. CONCLUSIONS

In summary, nanocomposite fibers for application in photonics have been realized by electrospinning, and their properties of emission and photon transport have been characterized. Various species of nanocomposite fibers can be considered as building blocks of novel optical materials, in which the spatial modulation of the refractive index, of the emission properties, and of the local light-scattering components can be tailored in a quite controlled way depending on the fabrication parameters and on the properties of the embedded nanoparticles. The analysis of the different contributions to optical losses highlights a main role played by the absorption and Rayleigh scattering by the inorganic fillers, as well as by the Rayleigh scattering by surface defects in nanocomposite light-emitting nanofibers. In perspective, these mechanisms and their control can be transferred to three-dimensional structures composed of electrospun filaments. The three-dimensional network made by electrospun polymer filaments is highly flexible, and they can be reliably bent. This might lead to designing new applications and devices in which the mutual distances among inorganic fillers can be varied in a controlled way, to possibly tune light-scattering properties, feedback mechanisms affecting radiation propagation and ultimately lasing through mechanical coupling with the complex photonic structures.

*Acknowledgments*. The research leading to these results has received funding from the European Research Council under the European Union's Seventh Framework Programme (FP/2007-2013)/ERC Grant Agreement n. 306357 (ERC Starting Grant "NANO-JETS"). The authors also acknowledge the Apulia Regional Projects 'Networks of Public Research Laboratories', WAFITECH (09) and M. I. T. T. (13), and D. Magrì for technical help.